\documentclass[12pt,preprint]{emulateapj}

\usepackage{psfig}
\usepackage{graphicx}
\usepackage{natbib}
\usepackage[english]{babel}  
\usepackage{txfonts}

\shorttitle{Shock-cloud interaction and particle acceleration in SN 1006}
\shortauthors{Miceli et al.}

\begin{document}


\title{Shock-cloud interaction and particle acceleration in the southwestern limb of SN 1006}

%% As in the title, use \\ to force line breakm s$^{-1}$.

\author{M. Miceli\altaffilmark{1}, F. Acero\altaffilmark{2}, G. Dubner\altaffilmark{3}, A. Decourchelle\altaffilmark{4}, S. Orlando\altaffilmark{1} and F. Bocchino\altaffilmark{1}}
%\affil{Astronomy Department, University of California,
%    Berkeley, CA 94720}

%\author{C. D. Biemesderfer\altaffilmark{4,5}}
%\affil{National Optical Astronomy Observatories, Tucson, AZ 85719}
%\email{aastex-help@aas.org}

%\and

%\author{R. J. Hanisch\altaffilmark{5}}
%\affil{Space Telescope Science Institute, Baltimore, MD 21218}


\altaffiltext{1}{INAF-Osservatorio Astronomico di Palermo, Piazza del Parlamento 1, 90134 Palermo, Italy \email{miceli@astropa.unipa.it}}
\altaffiltext{2}{ORAU/NASA Goddard Space Flight Center, Astrophysics Science Division, Code 661, Greenbelt, MD 20771}
\altaffiltext{3}{Instituto de Astronom\'ia y F\'isica del Espacio (IAFE), UBA-CONICET, CC 67, Suc. 28, 1428 Buenos Aires, Argentina}
%\altaffiltext{4}{Service d'Astrophysique/IRFU/DSM, CEA Saclay, Gif-sur-Yvette, France}
\altaffiltext{4}{Laboratoire AIM-Paris-Saclay, CEA/DSM/Irfu - CNRS - Universit\'e Paris Diderot, CE-Saclay, 91191 Gif-sur-Yvette, France}


\begin{abstract}
The supernova remnant SN~1006 is a powerful source of high-energy particles and evolves in a relatively tenuous and uniform environment, though interacting with an atomic cloud in its northwestern limb.
The X-ray image of SN 1006 reveals an indentation in the southwestern part of the shock front and the HI maps show an isolated  cloud (southwestern cloud) having the same velocity as the northwestern cloud and whose morphology fits perfectly in the indentation. 
We performed spatially resolved spectral analysis of a set of small regions in the southwestern nonthermal limb and studied the deep X-ray spectra obtained within the \emph{XMM-Newton} SN~1006 Large Program. We also analyzed archive HI data, obtained combining single dish and interferometric observations. We found that the best-fit value of the $N_H$ derived from the X-ray spectra significantly increases in regions corresponding to the southwestern cloud, while the cutoff energy of the synchrotron emission decreases. The amount of the $N_H$ variations corresponds perfectly with the HI column density of the southwestern cloud, as measured from the radio data. The decrease in the cutoff energy at the indentation clearly reveals that the back side of the cloud is actually interacting with the remnant. The southwestern limb therefore presents a unique combination of efficient particle acceleration and high ambient density, thus being the most promising region for $\gamma-$ray hadronic emission in SN 1006. We 
estimate that such emission will be detectable with the $Fermi$ telescope within a few years. 
%The presence of a dense environment near a region where efficient particle acceleration is at work makes the southwestern limb a promising source of $\gamma-$ray hadronic emission. We estimate that such emission will be detectable with the $Fermi$ telescope within a few years. 
\end{abstract}

\keywords{X-rays: ISM --- ISM: supernova remnants --- ISM: individual object: SN 1006 --- ISM: clouds --- acceleration of particles}

\section{Introduction}

The historical type Ia supernova remnant (SNR) SN~1006 is an ideal target to study the Fermi acceleration process in astrophysical shocks. It is a dynamically young remnant with shock velocity $v_s\sim 5000$ km$/$s (\citealt{kpl09,klp13,wwr14}), that evolves in a tenuous medium ($n_{0}\sim 0.035$ cm$^{-3}$ in the southeastern limb, see \citealt{mbd12}). It is spatially extended (radius $R\sim 15'$) and shows a rather simple bilateral non-thermal morphology. This allows us to study regions with highly efficient particle acceleration in the two opposed radio, X-ray, and $\gamma-$ray bright limbs and regions with less efficient particle acceleration in the northwestern and southeastern thermal limbs. 
Moreover, the X-ray spectrum of SN~1006 is characterized by a low interstellar absorption ($N_H\sim7\times10^{20}$ cm$^{-2}$, \citealt{dgg02}, hereafter D02).

The shape of the X-ray spectra extracted from the synchrotron limbs of SN 1006 reveals that the maximum energy achieved by electrons in the acceleration process is limited by their radiative losses (\citealt{mbd13}). The presence of electrons with cutoff energies of $\sim 10$ TeV suggests that also hadrons (that do not suffer radiative losses) can be efficiently accelerated up to ultrarelativistic energies at the shock front. \citet{mbd12} found that the shock compression ratio increases from 4 up to $\sim 6$ in regions of the southeastern rim that are closer to the nonthermal limbs. This can be naturally interpreted as a result of shock modification induced by hadron acceleration. 
More recently, \citet{nvh13} have revealed suprathermal hadrons in the northwestern limb of SN 1006, by studying the variations in the H$\alpha$ broad line widths and the broad-to-narrow line intensity ratios.

However, as shown by \citet{aaa10}, the TeV observations of SN 1006 obtained with the $HESS$ telescope cannot be uniquely  attributed to hadron emission (i.~e., proton-proton interactions with $\pi^0$ production and subsequent decay).
The data are consistent with a pure leptonic model (i.~e., inverse Compton from the accelerated electrons), in agreement with the morphology of the $\gamma-$ray emission of SN 1006, which also favors a leptonic origin \citep{pbm09}. On the other hand, \citet {aaa10} have shown that a mixed scenario that includes leptonic and hadronic components also provides a good fit to the gamma-ray data and according to the model by \citet{bkv12}, the hadronic and leptonic components in the $\gamma-$ray emissions are of comparable strength.
The basic model of hadronic $\gamma-$ray production requires particles accelerated up to multi-TeV energies and a target of sufficient density. Therefore, the tenuous environment around SN 1006 ($n_{0}<0.05$ cm$^{-3}$, see \citealt{abd07,mbd12}) does not favour the proton-proton interactions that yields the hadronic emission. A higher ambient density is observed in the northwestern rim, where the shock is slowed down by the interaction with dense material (hereafter northwestern cloud), producing a relatively bright and sharp H$_{\alpha}$ filament (e. g.,  \citealt{gwr02,wgl03,rks07}). However,  particle acceleration is not very efficient therein, as revealed by the lack of nonthermal X-ray emission.

Here we report the detection of a dense atomic cloud interacting with the southwestern synchrotron rim of SN 1006 (hereafter southwestern cloud), where efficient particle acceleration is at work.
The location of this cloud near an efficient site of particle acceleration makes the southwestern limb a very promising source of $\gamma-$ray hadronic emission.

\section{Results}
\label{results}
We here analyze the EPIC data of the \emph{XMM-Newton} Large Program of observations of SN~1006 (together with older \emph{XMM-Newton} archive observations). The data and the reduction process are described in detail in \citet{mbd12,mbd13}.
The interstellar environment around SN 1006 was studied by using the HI observations presented in D02. The radio data were obtained using the Australia Telescope Compact Array and combined with single dish data from Parkes telescope (we refer to D02 for further information). 
%The angular resolution is 4$^\prime$ $\times$ 3$^\prime$.7 and the noise level better than $\sim39$ mJy$/$beam (we refer to D02 for further information).

\subsection{Imaging analysis}

\begin{figure}[htb!]
 \centerline{\hbox{     
     \psfig{figure=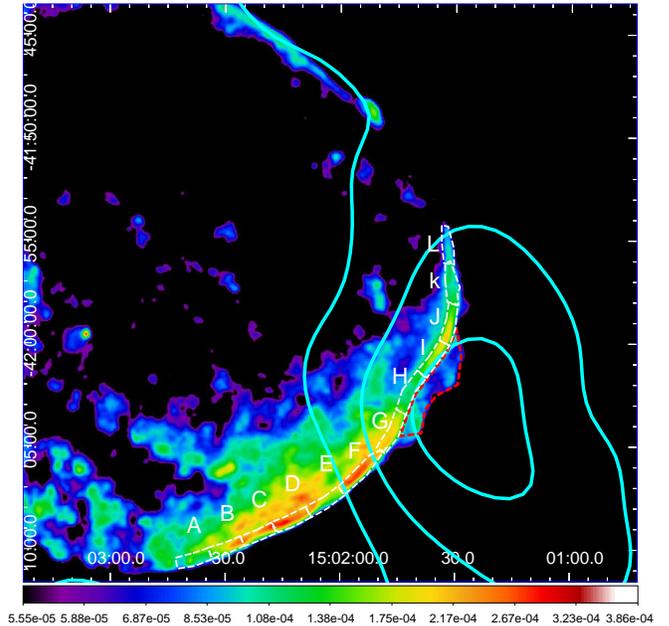,width=\columnwidth}}}
\caption{EPIC count-rate images (MOS and pn mosaic) of the southwestern part of SN~1006 in the $0.3-2$ keV band. The bin size is $2''$ and the image is background subtracted and adaptively smoothed to a signal-to-noise ratio of ten. The regions selected for the spectral analysis of the rim are superimposed: from $A$ to $L$ along the shock and the faint bulge ahead (red dashed line). The cyan curves indicate the contours of the column density (at $1.9\times 10^{20}$ cm$^{-2}$, $2.25\times 10^{20}$ cm$^{-2}$, and $2.7\times 10^{20}$ cm$^{-2}$) derived from the HI observations in the $[+5.8,+10.7]$ km s$^{-1}$ velocity range.}
\label{fig:X-radio}
\end{figure}

Figure \ref{fig:X-radio} shows the EPIC map of the southwestern quadrant of SN 1006 in the $0.3-2$ keV band. We notice a clear indentation in the shock front, corresponding to regions $G$, $H$, $I$, preceded by a faint "bulge" (red dashed region).

D02 carried out an extensive study of the atomic and molecular gas in the environs of SN~1006, concluding that at large scale the remnant evolves in a quite smooth, tenuous environment, as expected at high Galactic latitudes. At a more detailed scale, it can be noticed the presence of two HI structures that match peculiar features observed along the periphery of SN~1006. One of these structures is an elongated cloud abutting the flattened northwestern border of the SNR (i. e., the northwestern cloud), and the other is a cloud located to the southwest of SN~1006 (the southwestern cloud), whose border remarkably corresponds to the concavity of the shock front shown in Fig. \ref{fig:X-radio}.
Both these features peak at $v_{\rm LSR} \sim~+10$ km s$^{-1}$ and are clearly visible in Fig. 3b of D02. By assuming that the neutral gas is optically thin (as D02), we infer the atomic column density. 
Figure \ref{fig:X-radio} shows the contours of the column density estimated in the $[+5.8,+10.7]$ km s$^{-1}$ velocity range.
In spite of the excellent morphological agreement, D02 had proposed that another HI concentration at v$_{\rm LSR} \sim -$5  km s$^{-1}$ was a better candidate, although neither the gas at -5 km s$^{-1}$ nor the one at +10 km s$^{-1}$ had the expected systemic velocity for Galactic gas placed at $\sim$ 2 kpc. 
Revisiting the same data, it is now evident that the northwestern and southwestern clouds coincide with the northwestern and southwestern lobes of SN 1006.
In this picture, the elongated, flat structure of the northwestern cloud, naturally appears as the origin of the sharp H$_{\alpha}$ filaments observed all along the northwestern flank, and then this gas must be in the local environment of SN~1006. 
 
As noted before, these structures have anomalous kinematical velocities.
However, a velocity deviation from a Galactic circular rotation model is not surprising for gas placed almost $600$ pc above the Galactic plane, where the connection with the disk rotation is very unlikely. Also, in this part of the Milky Way, it is known that the disk bends towards more negative Galactic latitudes (\citealt{bur76}), increasing the z distance of these clouds from the Galactic arms. Moreover, by supposing that these structures do follow the circular rotation model, we obtain a very unlike outcome, since they would be giant clouds (larger than $\sim$ 100 pc), placed as far as $\sim$ 15 kpc in the halo of the Milky Way ($\sim3.7$ kpc above the Galactic plane), in perfect agreement with different portions of the border of SN~1006 only by chance. On the other hand, if we consider the hypothesis of cloud disruption as a consequence of the SNR expansion, a departure of $\sim$ 40 km s$^{-1}$ from the systemic velocity, would be naturally explained.
The X-ray spectral analysis described in the next section will further confirm that the southwestern cloud cannot be located farther than SN 1006.
By assuming that the southwestern cloud is at the same distance as SN~1006 (2.2 kpc), we estimate an atomic gas density of about $9-16$ cm$^{-3}$, by assuming an extension of the cloud along the line of sight of $8.5'-15.5'$ (i. e., the minimum$-$maximum angular extension in the plane of the sky).

\subsection{Spatially resolved spectral analysis}

We perform a spatially-resolved spectral analysis of the X-ray emission originating from regions $A$ to $L$ of Fig. \ref{fig:X-radio} to search for variations along the rim of column density and synchrotron cutoff energy $h\nu_{cut}$. We model the nonthermal emission with the loss-limited model developed by \citet{za07} (highly suited for SN 1006, as shown by \citealt{mbd13,mbd14}). Spectral analysis was performed in the $0.3-7.5$ keV band by using XSPEC V12.8.

Since we are interested in measuring the $N_H$ (the interstellar absorption is modelled through the TBABS model), it is important to correctly describe the soft thermal emission (mainly associated with O VII and O VIII line complexes). To model the thermal emission, we included a VPSHOCK component \citep{blr01}, with the same abundances as the ejecta component in region $e$ of \citet{mbd12}. We first focussed on region $A$ of Fig. \ref{fig:X-radio}, where the thermal emission is the highest, and determined the best-fit values of the temperature, $kT=0.6^{+0.2}_{-0.1}$ keV and ionization timescale $\tau=1.5^{+1.0}_{-0.6}\times10^{10}$ s cm$^{-3}$.
We then fixed these values when fitting the other spectra, and let the plasma emission measure as a free parameter. We refer to this procedure as approach 1. We verified that we do not find pronounced variations in the best-fit parameters, even if we model the thermal line emission with two narrow Gaussians (hereafter approach 2, see \citealt{mbd13}). In particular, the values of $N_H$ (and $h\nu_{cut}$) derived with the two approaches follow the same azimuthal trend and are all consistent within 2 sigmas, except for regions $A$, $B$, $C$, where the $N_H$ derived by adopting approach 2 are unrealistically low ($\sim5\times10^{20}$ cm$^{-2}$)\footnote{This is because this approach underestimates the low energy thermal continuum.}. 
Therefore, in the following, we report only the results obtained by adopting approach 1. All the spectra are very well described by our model and the reduced $\chi^2$ ranges between $1.00$ and $1.10$ (with $1100-2000$ dof).

\begin{figure}[htb!]
 \centerline{\hbox{     
     \psfig{figure=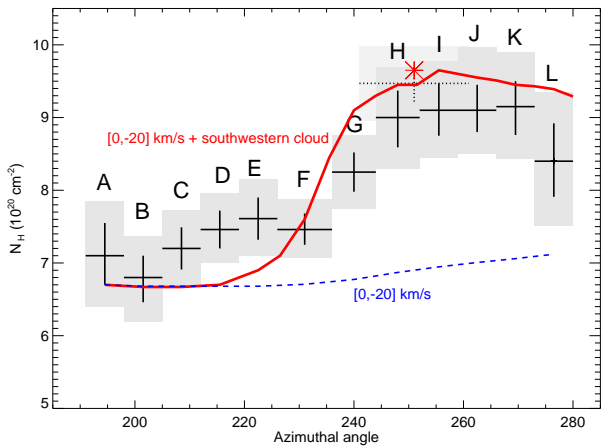,width=\columnwidth,angle=90}}}
\caption{Best fit values of the $N_H$ obtained from the X-ray spectral analysis of regions $A-L$ of Fig. \ref{fig:X-radio}. The dashed cross corresponds to the red dashed region ("the bulge") of Fig. \ref{fig:X-radio}. Error bars are shown at the $90\%$ (crosses) and $99\%$ (shaded areas) confidence levels. The blue curve shows the $N_H$ profile derived from the HI observations integrated in the $[0,-20]$ km s$^{-1}$ velocity range (D02). The red curve was obtained by adding to the blue curve the contribution of the southwestern cloud in the $[+5.8,+10.7]$ km s$^{-1}$ velocity range (the red star indicates the value at the "bulge").}
\label{fig:nh}
\end{figure}

Figure \ref{fig:nh} shows the azimuthal variations of the best-fit values of the $N_H$ for regions $A-L$. We observe a clear increase in the absorbing column in regions $G-L$, corresponding to the location of the southwestern cloud (see Fig. \ref{fig:X-radio}). In these regions, the $N_H$ is significantly higher than the characteristic value of $\sim 7\times10^{20}$ cm$^{-2}$, estimated by D02 integrating the HI radio emission in the $[0,-20]$ km s$^{-1}$ range (see dashed blue curve in Fig. \ref{fig:nh}, derived by sampling the $N_H$ values inferred by the HI observations). 
If we include also the contribution to the column density of the southwestern cloud (derived from the HI observations in the $[+5.8,+10.7]$ km s$^{-1}$ velocity range), we find that the azimuthal profile of the $N_H$ obtained from the radio observation is in amazing agreement with that derived by the X-ray spectral analysis. This result further confirms that the southwestern cloud lies at a distance compatible with that of SN 1006 (and foreground), and not at 15 kpc as estimated from a blindly applied circular rotation model.

\begin{figure}[htb!]
 \centerline{\hbox{     
     \psfig{figure=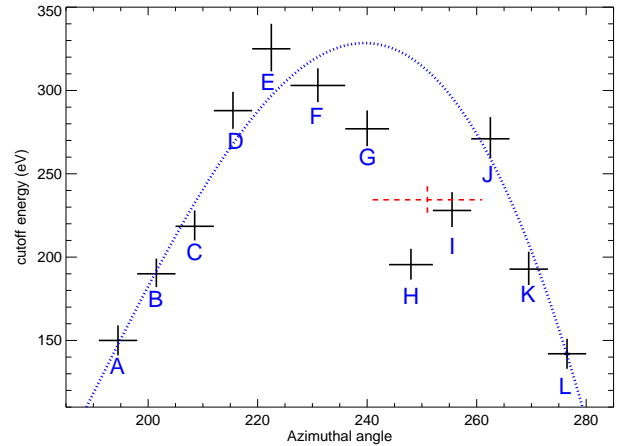,width=\columnwidth,angle=90}}}
\caption{Best fit values of the synchrotron cutoff energy obtained from the X-ray spectral analysis of regions $A-L$ of Fig. \ref{fig:X-radio}. The red cross corresponds to the red dashed region of Fig. \ref{fig:X-radio}. Error bars are shown at the $90\%$ confidence levels. The blue curve shows a 4th degree polynomial fit to all the points but $G,~H$, and $I$.}
\label{fig:break}
\end{figure}

The morphology of the southwestern cloud, perfectly corresponding to the concave shape of the southwestern shock front, strongly suggests that the back side of the cloud actually interacts with SN 1006. If this is the case, we expect the shock to be slowed down by this interaction. A direct consequence would be a net decrease in the cutoff frequency of the synchrotron emission in the interaction region, because in SN 1006 the maximum electron energy is limited by radiative losses and in the loss-limited scenario $h\nu_{cut}$ depends on $v_{s}^{2}$ \citep{za07}. 
Figure \ref{fig:break} shows the azimuthal profile of the best fit values of $h\nu_{cut}$ in regions $A-L$. Indeed, we find that the cutoff frequency abruptly drops down in regions $G,~H,~I$, i. e., exactly where the shock front is indented. Some hints for the presence of this dip in the azimuthal profile of the cutoff energy can be found in \citet{rbd04} and \citet{mbi09}, though in these papers the large regions selected for the spatially-resolved spectral analysis prevent a significant detection like the one presented here. We conclude that the bulk of the southwestern cloud lies in the foreground (as shown by the $N_H$ variations in Fig. \ref{fig:nh}) and part of it (i. e., regions $G,~H,~I$) is physically interacting with SN 1006. 

We searched for ISM thermal emission in region $H$ (where the ambient density is expected to be at its maximum), but the spectrum is synchrotron-dominated therein and very well described by our model (reduced $\chi^2=1.0$ with 1203 dof), so an additional thermal component is not statistically needed and the ISM temperature, $kT_{ISM}$, is unconstrained. However, we derived some temperature-dependent upper limits for the ISM post-shock density (deduced from the emission measure as in \citealt{mbd12}). We found $n_{ISM}<(12,~3,~0.3)$ cm$^{-3}$ for $kT_{ISM}=(0.05,~0.25,~2)$ keV, respectively. The values obtained for low temperatures (expected in case of propagation in a dense environment) are in agreement with the shock-cloud interaction scenario and similar to those derived from the HI analysis. The shock is probably interacting with the outer border of the cloud (where we expect lower densities), since the amount of the indentation indicates that the interaction is recent and the shock has not entered yet 
within the bulk of the cloud. We notice that the $H_\alpha$ emission in the southwhestern limb is quite faint and noisy (\citealt{wgl03}) and the lack of a clear $H_\alpha$ filament can be due to a heavily ionized preshock medium and$/$or to an unfavorable (not edge on) orientation of the emitting sheet. 

We obtain a heuristic estimate of the value that $h\nu_{cut}$ would have had in the case of no interaction by performing a 4th degree polynomial fit to all the points but  $G,~H,~I$ (the best fit curve is shown in Fig. \ref{fig:break}). We verified that this simple (unphysical) polynomial form also provides a good description of the  $h\nu_{cut}$ profile in the northeastern limb. The cutoff frequency in region $H$ is reduced by a factor $f\sim 1.7$. Since $h\nu_{cut}\propto v_{s}^2\propto n_{ISM}^{-1}$, one may suppose that in region $H$ the ambient density is higher by the same factor $f$ and therefore is slightly lower than $0.1$ cm$^{-3}$ (assuming $n_{ISM}\sim0.03-0.05$ cm$^{-3}$ elsewhere). This value is much smaller than the average density of the southwestern cloud estimated by the HI data and seems insufficient to produce the observed indentation in the shock. However, we notice that other (non-interacting) parts of the shell, whose projected location falls within region $H$, may contribute to the 
observed synchrotron emission, thus producing a biased overestimation of $h\nu_{cut}$ (relatively high $h\nu_{cut}$ have been observed well inside the shell by \citealt{rbd04}) and underestimation of $n_{ISM}$. In any case, in the concave region of the southwestern shock front, $h\nu_{cut}$ is significantly lower than in adjacent region, in agreement with what we expect in case of shock-cloud interaction.

The red dashed region in Fig. \ref{fig:X-radio} also shows low values of $h\nu_{cut}$ and high $N_H$. It may be associated with a non-interacting region located behind the cloud, or to particles leaving the remnant and diffusing inside the cloud. Further investigation are necessary to ascertain the origin of this "bulge"

\section{Discussion and conclusions}

We have found different evidences for a shock-cloud interaction in the southwestern limb of SN 1006: i) the southwestern shock presents a sharp indentation in the X-rays (as well as in the radio band, see Fig. 3 in \citealt{pdc09}); ii) The indentation corresponds to the position of an HI cloud; iii) this cloud has the same velocity as the northwestern cloud, which is interacting with SN 1006; iv) the variations of the $N_H$ derived from the X-ray spectra show that the southwestern cloud lies in the foreground; v) in the indentation region the synchrotron cutoff energy is significantly lower than in adjacent regions.
The southwestern limb, characterized by a highly efficient particle acceleration, therefore presents a unique combination of efficient particle acceleration and high target density, thus being the most promising region for $\gamma-$ray hadronic emission in SN 1006.

\begin{figure}[t!]
 \centerline{\hbox{     
     \psfig{figure=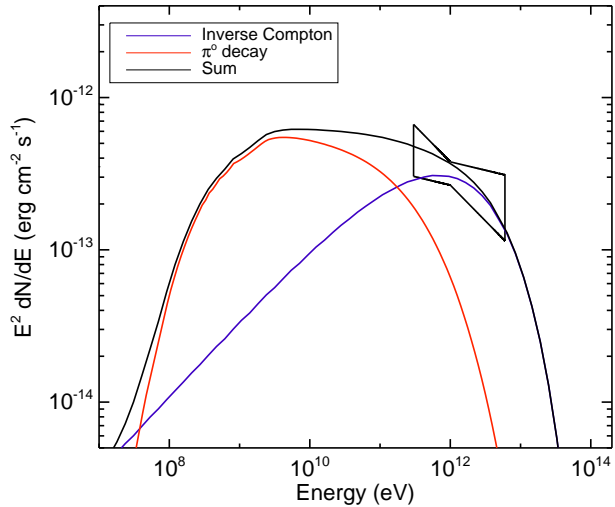,width=\columnwidth}}}
\caption{Gamma-ray spectrum of the southwestern limb of SN 1006 observed with $HESS$ \citep{aaa10}.
We superimpose our spectral model and highlight the hadronic contribution of the shock$/$cloud region (red) and the leptonic (blue) emission from the whole limb.}
\label{fig:gamma}
\end{figure}

We estimate the expected hadronic $\gamma-$ray emission from the southwestern limb of SN 1006 by adopting the phenomenological model described in \citet{aaa10}. The $\pi^0$ production from the shock$/$cloud region is calculated by following \citet{kab06} and the particle distribution is assumed to follow a power law ($\Gamma$=2.0 for electrons and hadrons) with an exponential cutoff, from which we also calculate the leptonic emission (synchrotron, bremsstrahlung, and IC scattering).
The fundamental parameters of our model are the particle total energy ($W_{\rm e, p}$), the particle distribution cutoff energy ($E^{\rm c}_{\rm e,p}$), and the ambient density. Considering that the shock is probably not yet interacting with the bulk of the cloud, we assume a downstream density of 10 cm$^{-3}$ (i. e., a pre-shock density smaller by at least a factor of 4-6), in agreement with the HI data analysis results and the upper limits provided by the X-ray spectral analysis (Sect. \ref{results}). 
%We set the cloud density to 10 cm$^{-3}$, in agreement with our results of the HI data analysis (Sect. \ref{results}). 
We assume that the total hadronic energy is $10\%$ of a canonical explosion energy of $10^{51}$ erg. This value was derived by considering that (near the nonthermal limbs) the shock compression ratio is $r\sim 6$ (\citealt{mbd12}) and the model by \citet{vyh10} (Eq. 15, therein, for Mach number $M>>1$) shows that this $r$ corresponds to a $10\%$ of the available energy transferred to cosmic rays. Also, we verified that the MHD models of SN 1006 performed by \citet{obm12} (in particular, their best-fit models EX-C3.5-D2-QPAR-G1.3 and PL-C3.5-D2-QPAR-G1.3, both including shock modification) provide a similar estimate of the hadronic energy drain ($\sim9\%$). To derive the hadronic emission from the southwestern cloud, we finally scale the hadronic energy by the proper geometric factor $1/128$ (obtained by considering the solid angle corresponding to the southwestern cloud). We assume that the bulk of the TeV $\gamma-$ray emission from the whole southwestern limb is leptonic (fitted parameters W$_{\rm e}= 5\times 10^{46}$ ergs and $E^{\rm c}_{\rm e} = 15$ TeV) and we derive the best-fit value $E^{c}_{p}\sim3$ TeV for the shock/cloud hadronic component (in the interaction region).

Figure \ref{fig:gamma} shows the comparison between our model of the $\gamma-$ray spectrum (hadronic and leptonic contributions are highlighted) and the $HESS$ spectral information of the southwestern limb \citep{aaa10}. We predict that the hadronic emission from the southwestern cloud shows a marked peak at $\sim 10$ GeV, reaching a flux $> 5\times10^{-13}$ erg cm$^{-2}$ s$^{-1}$ in the $3-30$ GeV band. This value is slightly below the current sensitivity achieved by $Fermi/LAT$, that corresponds to less than $8\times10^{-13}$ erg cm$^{-2}$ s$^{-1}$, in the $3-30$ GeV band (see Acero et al. in preparation for further details). We verified that even in the unlikely scenario where the whole HESS spectrum from the southwestern limb has a hadronic origin, the expected flux in the 3-30 GeV does not change significantly, only decreasing to $\sim4\times10^{-13}$ erg cm$^{-2}$ s$^{-1}$ (the proton cutoff energy increases to 60 TeV).

We point out that the $Fermi/LAT$ sensitivity improves rapidly with time. Moreover, the upcoming Fermi-LAT event reconstruction technique (Pass-8) will offer a marked increase in the effective area, better angular resolution, and lower background contamination (\citealt{aab13}). We conclude that the hadronic emission from the southwestern limb of SN~1006 (if any) will be detectable within a few years.
In addition, the Cherenkov Telescope Array (CTA, \citealt{actis11,acharya13}) will be able to perform spatially-resolved spectroscopy to reveal possible signatures of hadronic emission in the interacting region.

%% The \notetoeditor{TEXT} command allows the author to communicate
%% information to the copy editor.  This information will appear as a
%% footnote on the printed copy for the manuscript style file.  Nothing will
%% appear on the printed copy if the preprint or
%% preprint2 style files are used.


%% Included in this acknowledgments section are examples of the
%% AASTeX hypertext markup commands. Use \url without the optional [HREF]
%% argument when you want to print the url directly in the text. Otherwise,
%% use either \url or \anchor, with the HREF as the first argument and the
%% text to be printed in the second.

\acknowledgments
We thank the anonymous referee for their important comments and suggestions.
This paper was partially funded by the ASI-INAF contract I$/$009$/$10$/$0. GD is funded by CONICET and ANPCYT (Argentina) grants. AD acknowledges support from CNES.

%% in parentheses, after the keyword, but they will not be verified.

\bibliographystyle{apj}
%\bibliography{references}

\begin{thebibliography}{}
\expandafter\ifx\csname natexlab\endcsname\relax\def\natexlab#1{#1}\fi

\bibitem[{{Acero} {et~al.}(2007){Acero}, {Ballet}, \& {Decourchelle}}]{abd07}
{Acero}, F., {Ballet}, J., \& {Decourchelle}, A. 2007, \aap, 475, 883

\bibitem[{{Acero} {et~al.}(2010){Acero}, {Aharonian}, {Akhperjanian}, {Anton},
  {Barres de Almeida}, {Bazer-Bachi}, {Becherini}, {Behera}, {Beilicke},
  {Bernl{\"o}hr}, {Bochow}, {Boisson}, {Bolmont}, {Borrel}, {Brucker}, {Brun},
  {Brun}, {B{\"u}hler}, {Bulik}, {B{\"u}sching}, {Boutelier}, {Chadwick},
  {Charbonnier}, {Chaves}, {Cheesebrough}, {Conrad}, {Chounet}, {Clapson},
  {Coignet}, {Dalton}, {Daniel}, {Davids}, {Degrange}, {Deil}, {Dickinson},
  {Djannati-Ata{\"i}}, {Domainko}, {O'C.~Drury}, {Dubois}, {Dubus}, {Dyks},
  {Dyrda}, {Egberts}, {Eger}, {Espigat}, {Fallon}, {Farnier}, {Fegan},
  {Feinstein}, {Fiasson}, {F{\"o}rster}, {Fontaine}, {F{\"u}{\ss}ling},
  {Gabici}, {Gallant}, {G{\'e}rard}, {Gerbig}, {Giebels}, {Glicenstein},
  {Gl{\"u}ck}, {Goret}, {G{\"o}ring}, {Hauser}, {Hauser}, {Heinz},
  {Heinzelmann}, {Henri}, {Hermann}, {Hinton}, {Hoffmann}, {Hofmann},
  {Hofverberg}, {Holleran}, {Hoppe}, {Horns}, {Jacholkowska}, {de Jager},
  {Jahn}, {Jung}, {Katarzy{\'n}ski}, {Katz}, {Kaufmann}, {Kerschhaggl},
  {Khangulyan}, {Kh{\'e}lifi}, {Keogh}, {Klochkov}, {Klu{\'z}niak}, {Kneiske},
  {Komin}, {Kosack}, {Kossakowski}, {Lamanna}, {Lemoine-Goumard}, {Lenain},
  {Lohse}, {Marandon}, {Marcowith}, {Masbou}, {Maurin}, {McComb}, {Medina},
  {M{\'e}hault}, {Moderski}, {Moulin}, {Naumann-Godo}, {de Naurois}, {Nedbal},
  {Nekrassov}, {Nicholas}, {Niemiec}, {Nolan}, {Ohm}, {Olive}, {de O{\~n}a
  Wilhelmi}, {Orford}, {Ostrowski}, {Panter}, {Paz Arribas}, {Pedaletti},
  {Pelletier}, {Petrucci}, {Pita}, {P{\"u}hlhofer}, {Punch}, {Quirrenbach},
  {Raubenheimer}, {Raue}, {Rayner}, {Reimer}, {Renaud}, {de Los Reyes},
  {Rieger}, {Ripken}, {Rob}, {Rosier-Lees}, {Rowell}, {Rudak}, {Rulten},
  {Ruppel}, {Ryde}, {Sahakian}, {Santangelo}, {Schlickeiser}, {Sch{\"o}ck},
  {Sch{\"o}nwald}, {Schwanke}, {Schwarzburg}, {Schwemmer}, {Shalchi}, {Sushch},
  {Sikora}, {Skilton}, {Sol}, {Stawarz}, {Steenkamp}, {Stegmann}, {Stinzing},
  {Superina}, {Szostek}, {Tam}, {Tavernet}, {Terrier}, {Tibolla}, {Tluczykont},
  {van Eldik}, {Vasileiadis}, {Venter}, {Venter}, {Vialle}, {Vincent}, {Vink},
  {Vivier}, {V{\"o}lk}, {Volpe}, {Vorobiov}, {Wagner}, {Ward}, {Zdziarski},
  {Zech}, \& {H.E.S.S.~Collaboration}}]{aaa10}
{Acero}, F., {Aharonian}, F., {Akhperjanian}, A.~G., {et~al.} 2010, \aap, 516,
  A62

\bibitem[{{Acharya} {et~al.}(2013){Acharya}, {Actis}, {Aghajani}, {Agnetta},
  {Aguilar}, {Aharonian}, {Ajello}, {Akhperjanian}, {Alcubierre},
  {Aleksi{\'c}}, \& et~al.}]{acharya13}
{Acharya}, B.~S., {Actis}, M., {Aghajani}, T., {et~al.} 2013, Astroparticle
  Physics, 43, 3

\bibitem[{{Actis} {et~al.}(2011){Actis}, {Agnetta}, {Aharonian},
  {Akhperjanian}, {Aleksi{\'c}}, {Aliu}, {Allan}, {Allekotte}, {Antico},
  {Antonelli}, \& et~al.}]{actis11}
{Actis}, M., {Agnetta}, G., {Aharonian}, F., {et~al.} 2011, Experimental
  Astronomy, 32, 193

\bibitem[{{Atwood} {et~al.}(2013){Atwood}, {Albert}, {Baldini}, {Tinivella},
  {Bregeon}, {Pesce-Rollins}, {Sgr{\`o}}, {Bruel}, {Charles}, {Drlica-Wagner},
  {Franckowiak}, {Jogler}, {Rochester}, {Usher}, {Wood}, {Cohen-Tanugi}, \&
  {S.~Zimmer for the Fermi-LAT Collaboration}}]{aab13}
{Atwood}, W., {Albert}, A., {Baldini}, L., {et~al.} 2013, ArXiv e-prints,
  arXiv:1303.3514

\bibitem[{{Berezhko} {et~al.}(2012){Berezhko}, {Ksenofontov}, \&
  {V{\"o}lk}}]{bkv12}
{Berezhko}, E.~G., {Ksenofontov}, L.~T., \& {V{\"o}lk}, H.~J. 2012, \apj, 759,
  12

\bibitem[{{Borkowski} {et~al.}(2001){Borkowski}, {Lyerly}, \&
  {Reynolds}}]{blr01}
{Borkowski}, K.~J., {Lyerly}, W.~J., \& {Reynolds}, S.~P. 2001, \apj, 548, 820

\bibitem[{{Burton}(1976)}]{bur76}
{Burton}, W.~B. 1976, \araa, 14, 275

\bibitem[{{Dubner} {et~al.}(2002){Dubner}, {Giacani}, {Goss}, {Green}, \&
  {Nyman}}]{dgg02}
{Dubner}, G.~M., {Giacani}, E.~B., {Goss}, W.~M., {Green}, A.~J., \& {Nyman},
  L.-{\AA}. 2002, \aap, 387, 1047

\bibitem[{{Ghavamian} {et~al.}(2002){Ghavamian}, {Winkler}, {Raymond}, \&
  {Long}}]{gwr02}
{Ghavamian}, P., {Winkler}, P.~F., {Raymond}, J.~C., \& {Long}, K.~S. 2002,
  \apj, 572, 888

\bibitem[{{Katsuda} {et~al.}(2013){Katsuda}, {Long}, {Petre}, {Reynolds},
  {Williams}, \& {Winkler}}]{klp13}
{Katsuda}, S., {Long}, K.~S., {Petre}, R., {et~al.} 2013, \apj, 763, 85

\bibitem[{{Katsuda} {et~al.}(2009){Katsuda}, {Petre}, {Long}, {Reynolds},
  {Winkler}, {Mori}, \& {Tsunemi}}]{kpl09}
{Katsuda}, S., {Petre}, R., {Long}, K.~S., {et~al.} 2009, \apj, 692, L105

\bibitem[{{Kelner} {et~al.}(2006){Kelner}, {Aharonian}, \& {Bugayov}}]{kab06}
{Kelner}, S.~R., {Aharonian}, F.~A., \& {Bugayov}, V.~V. 2006, \prd, 74, 034018

\bibitem[{{Miceli} {et~al.}(2012){Miceli}, {Bocchino}, {Decourchelle},
  {Maurin}, {Vink}, {Orlando}, {Reale}, \& {Broersen}}]{mbd12}
{Miceli}, M., {Bocchino}, F., {Decourchelle}, A., {et~al.} 2012, \aap, 546, A66

\bibitem[{{Miceli} {et~al.}(2013{\natexlab{a}}){Miceli}, {Bocchino},
  {Decourchelle}, {Vink}, {Broersen}, \& {Orlando}}]{mbd14}
---. 2013{\natexlab{a}}, ArXiv e-prints, arXiv:1309.1414

\bibitem[{{Miceli} {et~al.}(2013{\natexlab{b}}){Miceli}, {Bocchino},
  {Decourchelle}, {Vink}, {Broersen}, \& {Orlando}}]{mbd13}
---. 2013{\natexlab{b}}, \aap, 556, A80

\bibitem[{{Miceli} {et~al.}(2009){Miceli}, {Bocchino}, {Iakubovskyi},
  {Orlando}, {Telezhinsky}, {Kirsch}, {Petruk}, {Dubner}, \&
  {Castelletti}}]{mbi09}
{Miceli}, M., {Bocchino}, F., {Iakubovskyi}, D., {et~al.} 2009, \aap, 501, 239

\bibitem[{{Nikoli{\'c}} {et~al.}(2013){Nikoli{\'c}}, {van de Ven}, {Heng},
  {Kupko}, {Husemann}, {Raymond}, {Hughes}, \& {Falc{\'o}n-Barroso}}]{nvh13}
{Nikoli{\'c}}, S., {van de Ven}, G., {Heng}, K., {et~al.} 2013, Science, 340,
  45

\bibitem[{{Orlando} {et~al.}(2012){Orlando}, {Bocchino}, {Miceli}, {Petruk}, \&
  {Pumo}}]{obm12}
{Orlando}, S., {Bocchino}, F., {Miceli}, M., {Petruk}, O., \& {Pumo}, M.~L.
  2012, \apj, 749, 156

\bibitem[{{Petruk} {et~al.}(2009{\natexlab{a}}){Petruk}, {Bocchino}, {Miceli},
  {Dubner}, {Castelletti}, {Orlando}, {Iakubovskyi}, \& {Telezhinsky}}]{pbm09}
{Petruk}, O., {Bocchino}, F., {Miceli}, M., {et~al.} 2009{\natexlab{a}},
  \mnras, 399, 157

\bibitem[{{Petruk} {et~al.}(2009{\natexlab{b}}){Petruk}, {Dubner},
  {Castelletti}, {Bocchino}, {Iakubovskyi}, {Kirsch}, {Miceli}, {Orlando}, \&
  {Telezhinsky}}]{pdc09}
{Petruk}, O., {Dubner}, G., {Castelletti}, G., {et~al.} 2009{\natexlab{b}},
  \mnras, 393, 1034

\bibitem[{{Raymond} {et~al.}(2007){Raymond}, {Korreck}, {Sedlacek}, {Blair},
  {Ghavamian}, \& {Sankrit}}]{rks07}
{Raymond}, J.~C., {Korreck}, K.~E., {Sedlacek}, Q.~C., {et~al.} 2007, \apj,
  659, 1257

\bibitem[{{Rothenflug} {et~al.}(2004){Rothenflug}, {Ballet}, {Dubner},
  {Giacani}, {Decourchelle}, \& {Ferrando}}]{rbd04}
{Rothenflug}, R., {Ballet}, J., {Dubner}, G., {et~al.} 2004, \aap, 425, 121

\bibitem[{{Vink} {et~al.}(2010){Vink}, {Yamazaki}, {Helder}, \&
  {Schure}}]{vyh10}
{Vink}, J., {Yamazaki}, R., {Helder}, E.~A., \& {Schure}, K.~M. 2010, \apj,
  722, 1727

\bibitem[{{Winkler} {et~al.}(2003){Winkler}, {Gupta}, \& {Long}}]{wgl03}
{Winkler}, P.~F., {Gupta}, G., \& {Long}, K.~S. 2003, \apj, 585, 324

\bibitem[{{Winkler} {et~al.}(2014){Winkler}, {Williams}, {Reynolds}, {Petre},
  {Long}, {Katsuda}, \& {Hwang}}]{wwr14}
{Winkler}, P.~F., {Williams}, B.~J., {Reynolds}, S.~P., {et~al.} 2014, \apj,
  781, 65

\bibitem[{{Zirakashvili} \& {Aharonian}(2007)}]{za07}
{Zirakashvili}, V.~N., \& {Aharonian}, F. 2007, \aap, 465, 695

\end{thebibliography}




\end{document}